# Climates and clades: biased methods, biased results

Charles C. Davis[a], Hanno Schaefer[a,b], Brad R. Ruhfel[a,c],
Michael J. Donoghue[d], & Erika J. Edwards[d]

[a]*Department of Organismic and Evolutionary Biology, Harvard University, 22 Divinity Ave., Cambridge, MA 02138, USA*

[b]*Plant Biodiversity Research, Technische Universitaet Muenchen, D-85354 Freising, Germany*

[c] *Department of Biological Sciences, Eastern Kentucky University, Richmond, KY 40475, USA*

[d]*Department of Ecology and Evolutionary Biology, Yale University, P.O. Box 208106, New Haven, CT 06520, USA*

[d]*Department of Ecology and Evolutionary Biology, Brown University, Providence, RI 02912, USA*

**Correspondence** should be addressed to C.C.D (email: cdavis@oeb.harvard.edu)

Thuiller et al.[1] analyzed the consequences of anticipated climate change on plant, bird, and mammal phylogenetic diversity (PD) across Europe. They concluded that species loss will not be clade specific across the Tree of Life, and that there will not be an overall decline in PD across the whole of Europe. We applaud their attempt to integrate phylogenetic knowledge into scenarios of future extinction[2,3] but their analyses raise a series of concerns. We focus here on their analyses of plants.

First, their taxonomic sampling and phylogenetic methods have created biases that cast serious doubt on their conclusions. The plant species they sampled were drawn from the *Atlas Flora Europaea*[4], which presently includes a relatively small portion of the European Flora. As a result, nearly two thirds of their 1275 included species belong to a mere three clades: mustards (Brassicaceae, 24%), pinks (Caryophyllaceae, 22%), and buttercups + poppies (Ranunculales, 18%). This sampling is biased against woody plants, which are represented in their analyses by a small number of conifers, oaks (and relatives), and willows. Furthermore, enormous sectors of the plant Tree of Life that are well represented in Europe are entirely absent, such as monocotyledons (including grasses, sedges, lilies, and orchids), legumes, sunflowers, mints, umbels, and heaths (Fig. 1). Although little is known about the distribution of branch lengths in phylogenetic trees of whole clades or of regional floras, their skewed sampling yields what appears to be a highly unusual phylogenetic tree, with an overabundance of extremely short branches (Fig. 1). The fact that there are so few branches in the mid- to long-size range biases the predicted loss of PD to be small, and likely also explains why they failed to find higher extinction probabilities associated with longer branches. Phylogenetically random extinction on a tree with a more realistic range of branch lengths would likely yield a greater loss of PD than could ever be predicted using their tree.

We also doubt their analyses of phylogenetic signal in extinction risk. Thuiller et al.[1] used DNA sequence data from one exemplar of each of 378 genera to build a backbone tree; the remaining 896 species were placed into the relevant genera and resolved by simulating a Yule branching process. That is, nearly three quarters of the nodes in their 'high-resolution' phylogeny were randomly resolved. With such a shortage of phylogenetic information in their tree, it is not surprising that they found insignificant phylogenetic signal in their trait data.

Setting aside these critical analytical issues, we have additional concerns with their general approach. First, they equate extinction risk with the ability of a species to track climate and maintain its range size under the assumption that it will retain its current climate envelope. But we know that species respond differently to changes in climate. For example, some species shift their phenology more readily than others[5]. The relative ability to respond *in situ* to climate change will affect their estimates of 'suitable habitat', and we know that in Europe and the United States phenological response shows strong phylogenetic signal[6,7,8]. Their assumption of unlimited dispersal, which is unlikely in light of natural[9] and anthropogenic barriers in Europe, is equally problematical. Dispersal potential should also be analyzed in a phylogenetic context, since members of some clades are much more vagile than others. This is already evident in Europe where there has been northward movement of Mediterranean orchids, with tiny wind-blown seeds, as compared, for example, to legumes[10].

For all of these reasons we feel that no definitive conclusions can be drawn from the Thuiller et al.[1] study. Carefully crafted phylogenetic studies will certainly improve our ability to predict the impacts of climate change on biodiversity and devise better management plans. But the value of such studies, which have important societal implications, depends directly on the quality of the underlying phylogenetic and distributional data. Large-scale phyloinformatic analyses are exciting and promising, but we urge greater care in evaluating the biases that can confound such studies.

**Figure 1.** Generalized relationships of the plant genera of Germany[11] showing (in red) the limited sample of genera included by Thuiller et al.[1]. Density histograms of branch lengths for their phylogeny (in red; > 90% of the branch lengths are extremely short), compared to a recent well sampled community phylogeny of the flora of the Azores, Portugal[12] (in black, with ~ 800 species ), and a phylogeny of Poaceae[13] (grey, with 1,230 of ~12,000 species).

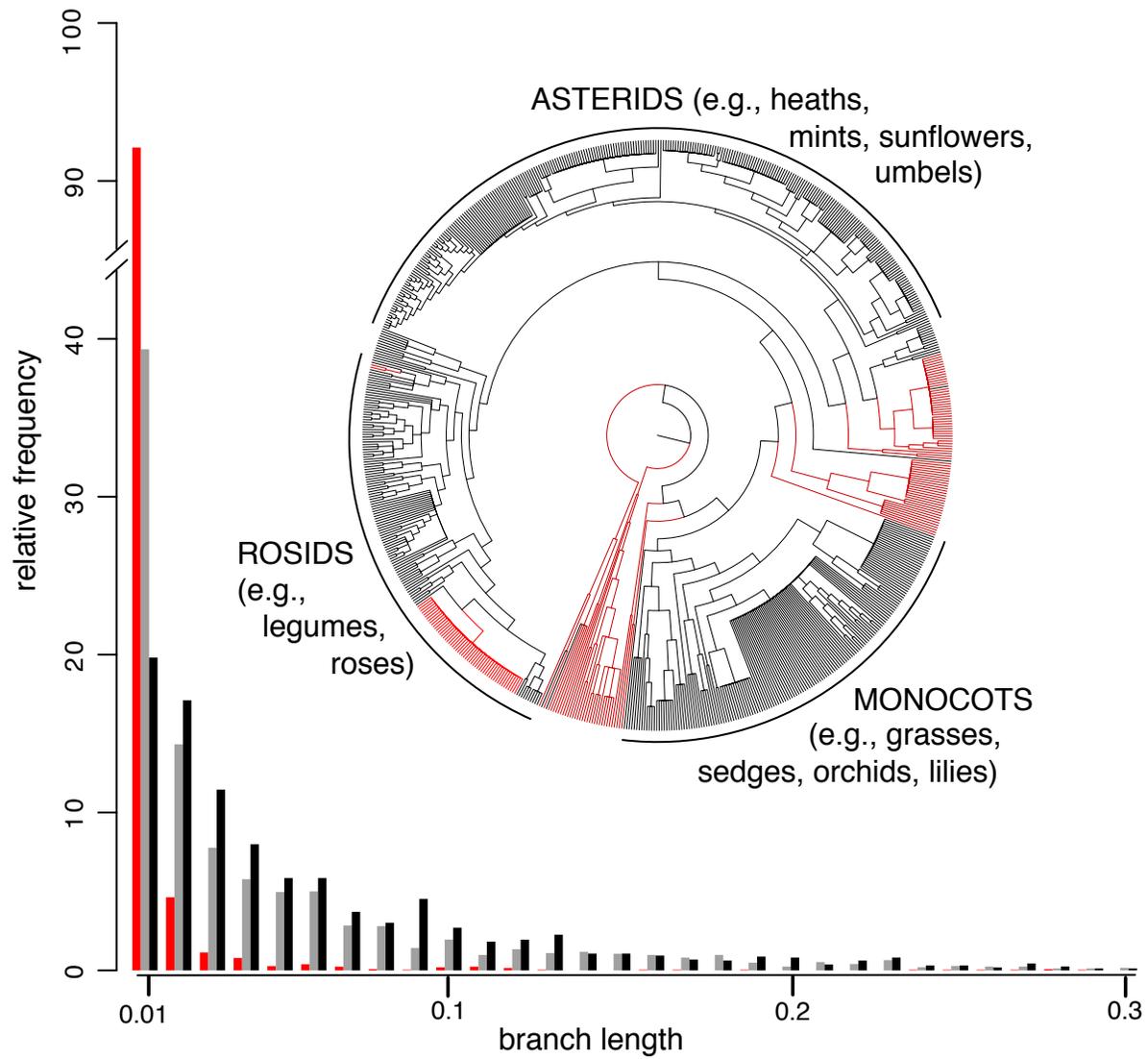